\documentclass[preprint,showpacs,preprintnumbers,amsmath,amssymb,nofootinbib]{revtex4}
\usepackage{booktabs}
\usepackage{mathrsfs}
\usepackage{epsfig}
\usepackage{graphicx}
\usepackage{dcolumn}
\usepackage{bm}
\usepackage{amsmath}
\usepackage{slashed}
\usepackage{multirow}
\usepackage{color}

\let\jnfont=\rm
\def\NPB#1,{{\jnfont Nucl.\ Phys.\ B }{\bf #1},}
\def\PLB#1,{{\jnfont Phys.\ Lett.\ B }{\bf #1},}
\def\EPJC#1,{{\jnfont Eur.\ Phys.\ Jour.\ C }{\bf #1},}
\def\PRD#1,{{\jnfont Phys.\ Rev.\ D }{\bf #1},}
\def\PRL#1,{{\jnfont Phys.\ Rev.\ Lett.\ }{\bf #1},}
\def\MPLA#1,{{\jnfont Mod.\ Phys.\ Lett.\ A }{\bf #1},}
\def\JPG#1,{{\jnfont J.\ Phys.\ G}{\bf #1},}
\def\CTP#1,{{\jnfont Commun.\ Theor.\ Phys.\ }{\bf #1},}
\def\ZPC#1,{{\jnfont Z.\ Phys.\ C }{\bf #1},}
\def\JHEP#1,{{\jnfont JHEP \ }{\bf #1},}
\def\Rv{\not{\hbox{\kern-1pt $R$}}}
\def\p{\not{\hbox{\kern-3pt $p$}}}

\newcommand{\bea}{\begin{eqnarray}}
\newcommand{\eea}{\end{eqnarray}}

\newcommand{\bcen}{\begin{center}}
\newcommand{\ecen}{\end{center}}

\newcommand{\ee}{e^+e^-}

\newcommand{\beq}{\begin{eqnarray}}
\newcommand{\eeq}{\end{eqnarray}}

\def\be{\begin{equation}}
\def\ee{\end{equation}}
\def\bea{\begin{array}}
\def\eea{\end{array}}
\def\beqa{\begin{eqnarray}}
\def\eeqa{\end{eqnarray}}
\def\beqas{\begin{eqnarray*}}
\def\eeqas{\end{eqnarray*}}

\def\bp{\begin{picture}}
\def\ep{\end{picture}}
\def\bc{\begin{center}}
\def\ec{\end{center}}
\def\bfig{\begin{figure}}
\def\efig{\end{figure}}

\def\bit{\begin{itemize}}
\def\eit{\end{itemize}}

\def\[{\left[}
\def\]{\right]}
\def\({\left(}
\def\){\right)}

\def\..{\left.}
\def\.{\right.}

\def\ep{\epsilon}

\def\t1{\tilde{t_1}}

\begin{document}

\title{750 GeV Diphoton Resonance in a Vector-like Extension of Hill Model}
\author{ Ning Liu$^{1,2}$}
\author{ Wenyu Wang$^3$}
\author{ Mengchao Zhang$^4$}
\author{ Rui Zheng$^5$}
\affiliation{
$^1$ Institute of Theoretical Physics, Henan Normal University, Xinxiang 453007, China\\
$^2$ ARC Centre of Excellence for Particle Physics at the Terascale, School of Physics, The University of Sydney, NSW 2006, Australia\\
$^3$ Institute of Theoretical Physics, College of Applied Science, Beijing University of Technology, Beijing 100124, China\\
$^4$ Key Laboratory of Theoretical Physics, Institute of Theoretical Physics, Chinese Academy of Science, Beijing 100190, China\\
$^5$ Department of Physics, University of California, Davis, CA 95616, USA
\vspace*{1.5cm} }

\begin{abstract}
In this paper, we study the recent 750 GeV diphoton excess in the Hill Model with vector-like fermions, in which the singlet-like Higgs boson is chosen as 750 GeV resonance and is mainly produced by the gluon fusion through vector-like top and bottom quarks. Meanwhile its diphoton decay rate is greatly enhanced by the vector-like lepton. Under the current experimental and theoretical constraints, we present the viable parameter space that fits the 750 GeV diphoton signal strength at 13 TeV LHC. We find that the heavier vector-like fermion masses are, the smaller mixing angle $\theta$ is required. The mixing angle of singlet and doublet Higgs bosons is constrained within $|\sin\theta| \lesssim 0.15$ in the condition of the perturbative Yukawa couplings. In the allowed parameter space, the 750 GeV diphoton cross section can be maximally enhanced to about 6 fb at 13 TeV LHC.
\end{abstract}

\maketitle

\section{INTRODUCTION}
ATLAS and CMS collaborations have reported an intriguing excess in the search for the resonance decaying into diphoton at 750 GeV, which correspond to a local (global) significance of $3.6 \sigma~(2.0 \sigma)$ and $2.6 \sigma~(1.2 \sigma)$, respectively \cite{atlas-diphoton,cms-diphoton}. The significance of ATLAS result slightly increases for a large width assumption ($\Gamma/m \sim 6\%$). While the CMS data favors a narrow width resonance.

After Moriond EW 2016, such an excess still persists and the kinematical distributions of the final states have been also presented by ATLAS. Since there are no other objects observed in the final states, the heavier resonance cascade decay and the heavy quark annihilation production of 750 GeV resonance seem disfavored. Besides, the ATLAS and CMS collaborations have searched for the high mass resonance in the $jj$ \cite{jj1}, $ZZ$ \cite{zz1,zz2}, $Z\gamma$ \cite{zabound}, $W^+W^-$ \cite{zz1,zz2}, $hh$\cite{hh1,hh2} and $t\bar{t}$ \cite{tt1,tt2} channels. But no significant excesses were reported. Combined with the 8 TeV diphoton data \cite{aa1,aa2}, the production rate of 750 GeV resonance is given by \cite{ex-6}
\begin{equation}
\sigma^{750}_{\gamma\gamma} = (4.6\pm 1.2) ~\rm{fb}~.\label{excess}
\end{equation}

So far, many new physics models for 750 GeV resonance have been proposed \cite{ex-1,ex-10,other}. From the view of naturalness, such a resonance may be the heavy singlet scalar in theories with extended Higgs sector, such as the CP-even singlet-like Higgs boson in NMSSM or the radion in Randall-Sundrum model. In general, the CP-even singlet scalar will naturally mix with the SM Higgs doublet, and thus couples to the SM particles via the mixing. Due to the constraint of LHC data, the mixing angle between the singlet and doublet should be small so that the diphoton production rate is highly suppressed. Thus, the additional new charged particles, in particular the vector-like fermions, are usually needed to enhance the diphoton production rate. However, if the new particles in the loops carry large quantum numbers and/or large multiplicity, the perturbativity of the SM gauge groups may break down below the Planck scale \cite{running-1,running-2}. This may indicate the new strong dynamics will appear at some high scale.

In this work, we will interpret the 750 GeV diphoton resonance in the Hill model with vector-like fermions. The singlet scalar is chosen as the 750 GeV resonance and strongly couples to diphoton through the vector-like charged fermions. Under the current experimental constraints, we find that the 750 GeV diphoton excess can be explained in this model. This paper is organized as follows. In Sec. \ref{sec2}, we describe the Hill model with vector-like fermions. In Sec. \ref{sec3} we present the numerical results and discussions. The conclusion is given in Sec. \ref{sec4}.

\section{Hill model with vector-like fermions}\label{sec2}
The starting point of our study is the Hill model~\cite{Hill:1987ea, Basso:2012nh}. In this model, the scalar Lagrangian reads
\begin{equation}\label{hilllagrangian}
\mathcal{L}_{\text{scalar}} = \left( D_\mu \Phi \right) ^\dagger \left( D^\mu \Phi \right) + \frac{1}{2} \left(\partial _\mu S\right) ^2 - \frac{\lambda _1}{2} \left( \Phi^\dagger \Phi - \frac{v^2}{2}\right) ^2 - \frac{\lambda _2}{2} \left( f_2 S - \Phi^\dagger \Phi \right) ^2.
\end{equation}
where $\Phi$ is the Higgs doublet field in SM and $S$ the extra singlet scalar field (Hill field) in Hill model.
This potential is minimized at $\left<\Phi\right>=v/\sqrt{2}$ and $\left< S\right> = v^2/(2 f_2)$.
After spontaneous symmetry breaking, the CP-even fields are defined as $\Phi^0 = (v+\phi)/\sqrt{2}$ and $S=s+\left< S\right>$. Due to the $\lambda_2$ term in Eq.~(\ref{hilllagrangian}), the two fields $\phi$ and $s$ will mix as
\begin{equation}
\begin{pmatrix}
\phi \\ s
\end{pmatrix}=
\begin{pmatrix}
c_\theta & -s_\theta \\
s_\theta & c_\theta
\end{pmatrix}
\begin{pmatrix}
h \\ H
\end{pmatrix}
\end{equation}
where we have used the shorthand notation $s_\theta=\sin\theta$ and $c_\theta=\cos\theta$ with $\theta$ the mixing angle of the two fields.
$h$ and $H$ are the two mass eigenstates and by convention $H$ is the heavier one. Their masses are given by
\begin{equation}\label{h_mas}
m_{h,H} = \frac{1}{2} \left( \lambda_2 f_2^2 + \lambda_3 v^2\right) \pm
\sqrt{\lambda_2^2 v^2 f_2^2 + \frac{1}{4}(\lambda_2 f_2^2 - \lambda_3 v^2)^2}\, ,
\end{equation}
where $\lambda_3 = \lambda_1 + \lambda_2$. In terms of the physical masses of $m_h$ and $m_H$, the mixing angle can be written as
\begin{equation}\label{h_ang}
c_\theta^2 = \frac{\lambda_2 f_2^2-m_h^2}{m_H^2-m_h^2}.
\end{equation}
Then the dimensionless parameters $\lambda_2$ and $\lambda_3$ can be given as,
\begin{align}
\lambda_2 &=\frac{s_\theta^2 c_\theta^2(m_H^2-m_h^2)^2}{\Big(m_H^2-s_\theta^2(m_H^2-m_h^2)\Big)v^2},\\
\lambda_3 &=\frac{m_h^2+s_\theta^2(m_H^2-m_h^2)}{v^2},\\
f_2 &= \frac{\Big(m_H^2-s_\theta^2(m_H^2-m_h^2)\Big)v}{c_\theta |s_\theta| (m_H^2-m_h^2)}.
\end{align}
In our study, we identify the lighter scalar $h$ as the 125 GeV SM Higgs boson and the heavier scalar $H$ as the 750 GeV diphoton resonance. Due to the mixing between $h$ and $H$, the heavy singlet-like scalar $H$ can decay to the SM dibosons, the fermion pair and Higgs pair. The current LHC measurements of the Higgs boson and the null results of searching for high mass resonance require the mixing angle $\sin\theta$ to be small. This will highly suppress the dominant production rate of $gg \to H$ and the decay width of $H \to \gamma\gamma$.

To solve this problem, we introduce the vector-like quarks $U^\prime_{L,R}$, $D^\prime_{L,R}$ and the vector-like lepton $L^\prime_{L,R}$\footnote{It should be noted that only $U^\prime$ or/and $D^\prime$ quark with the same electric charge as $t$ $(b)$} transforming as singlets under the electroweak $SU(2)_L$ gauge symmetry. Their SM quantum numbers are given by,
\beqa\label{quark}
U^\prime_L,U^\prime_R: (~3,~1,~2/3)~,~~~ D^\prime_L,D^\prime_R: (~3,~1,-1/3)~,~~~ L^\prime_L,L^\prime_R: (~1,~1,-1).
\eeqa
The couplings of the vector-like fermions with the Hill field are determined by,
\begin{equation}
\mathcal{L}_{\text{fermion}}=\sum_{F=U^\prime,D^\prime,L^\prime}\bar{F}(i\gamma^\mu D_\mu-m_{0,F}-y_{F} S) F
\end{equation}
After spontaneous symmetry breaking, the vector-like quark masses are $m_f=m_{0,F}+y_{F} \langle S \rangle$. Since the couplings of vector-like fermions to the Hill scalar are not proportional to their masses, we can separate the vector-like fermion masses $m_F$ from the Yukawa couplings  $y_{F}$. This feature can potentially enhance the effective couplings of $Hgg$ and $H\gamma\gamma$. Besides, we require $m_H < 2 m_{F}$ to kinematically forbid the decay channel $H \to F\bar{F}$, where $F=U^\prime,D^\prime,L^\prime$.

\section{Numerical Results and Discussions}\label{sec3}

\begin{figure}[ht]
\centering
\includegraphics[width=8cm]{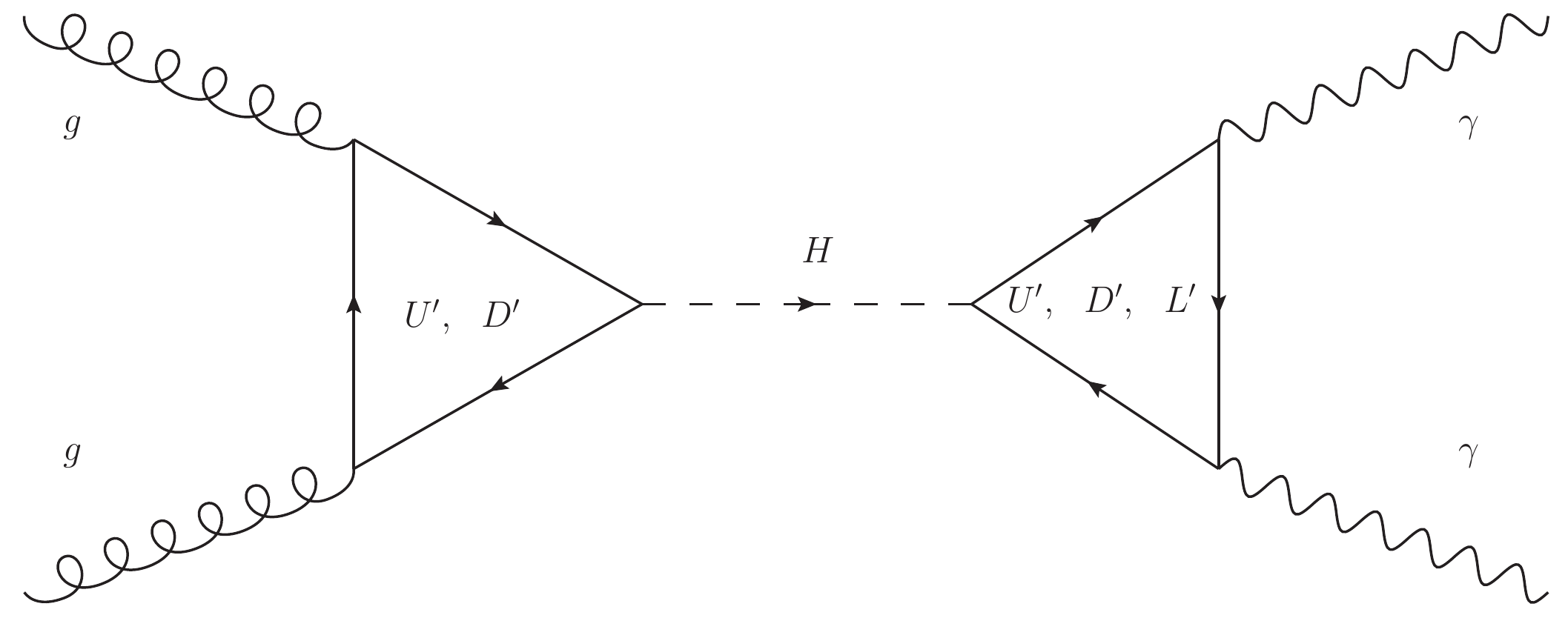}
\caption{Feynman diagrams for the process $gg \to H(750~{\rm GeV}) \to \gamma\gamma$.}
\label{feynman}
\end{figure}
In Fig.\ref{feynman}, we present the Feynman diagrams for the process $gg \to H(750~{\rm GeV}) \to \gamma\gamma$.
The gluon fusion production of $H$ is induced by the vector-like quarks $U^\prime$ and $D^\prime$, while the diphoton decay
of $H$ is induced by vector-like quarks and lepton $L^\prime$. We calculate the production cross section of $gg \to H$ at 13 TeV LHC by using the package \textsf{HIGLU}~\cite{higlu} with CTEQ6.6M PDFs~\cite{cteq6}. The renormalization and factorization scales are taken as $\mu_R=\mu_F=m_S/2$. We include the $K$-factor $(1+67\alpha_s/4\pi)$ ~\cite{qcdcorrection} in the calculation of the partial width of $H \to gg$. The SM parameters used in our numerical calculation are \cite{pdg}:
\begin{eqnarray}
  & \alpha_{ew} = 1/128 , \quad m_W = 80.385\mathrm{GeV} , \quad m_Z = 91.1876 \mathrm{GeV},  \quad m_t = 173.5 \mathrm{GeV}. &
\end{eqnarray}
The additional input parameters are mixing angle $\theta$, vector-like fermion masses $m_F$ and Yukawa couplings $y_F$, where $F=U^\prime,D^\prime,L^\prime$.

\begin{figure}[ht]
	\centering
	\includegraphics[width=10cm,height=6cm]{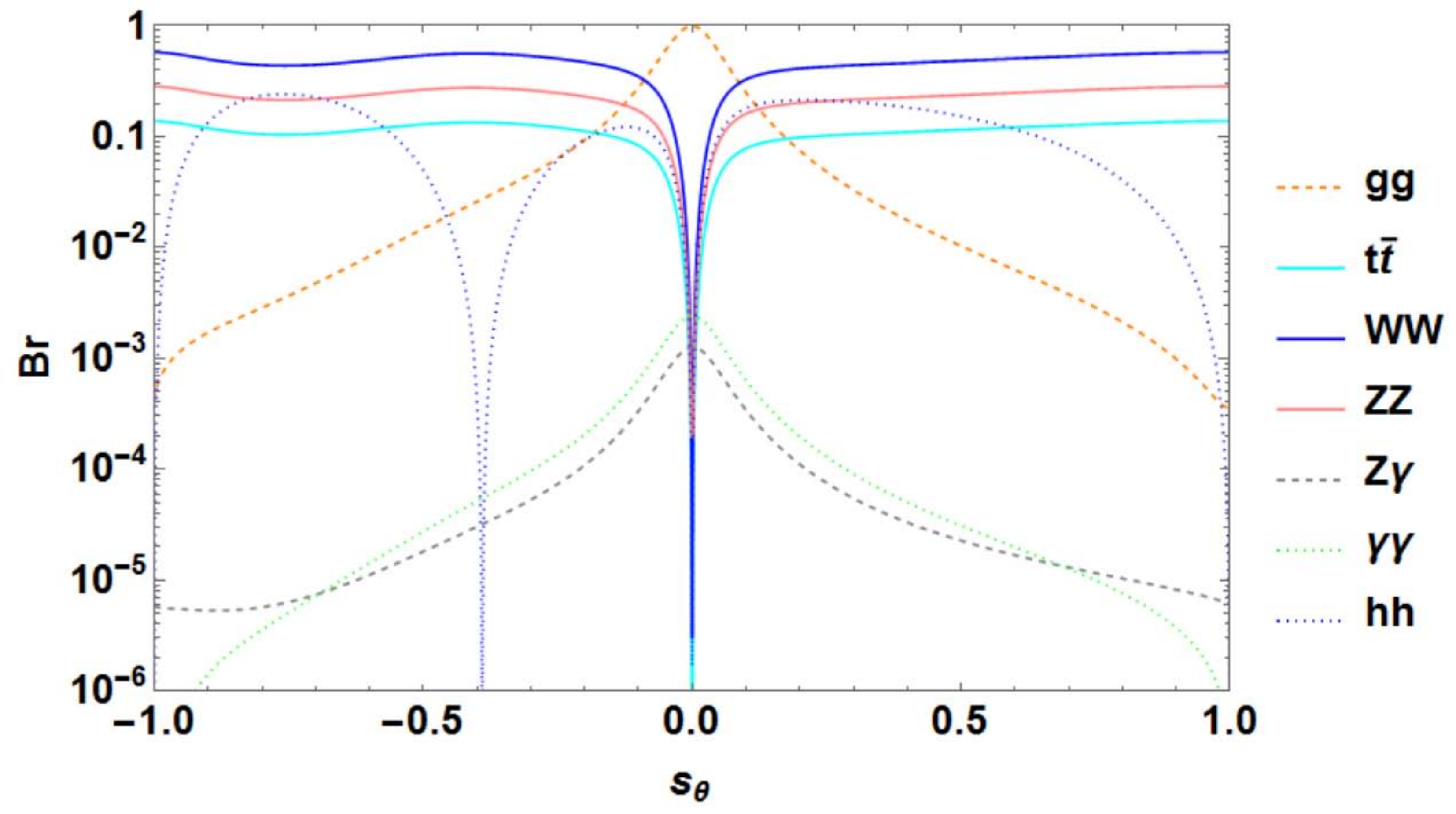}
\vspace{-0.5cm}
	\caption{Dependence of branching ratios of 750 GeV resonance $H$ on the mixing angle $\sin\theta$. We assume the universal mass $m_{U^\prime}=m_{D^\prime}=m_{L^\prime}=m_F=375$ GeV and the common yukawa couplings $y_{U^\prime}=y_{D^\prime}=y_{L^\prime}=y_F=2$. }
	\label{fig:sBr}
\end{figure}

In Fig.~\ref{fig:sBr}, we plot the dependence of the branching ratios of 750 GeV resonance $H$ on the mixing angle $\sin\theta$. For simplicity, we assume a common Yukawa couplings of vector-like fermions $y_{U^\prime}=y_{D^\prime}=y_{L^\prime}=y_F=2$ and a universal mass $m_{U^\prime}=m_{D^\prime}=m_{L^\prime}=m_F=375$ GeV. From Fig.~\ref{fig:sBr}, we can see that when $\sin\theta=0$, all the tree-level decay channels of $H$ will vanish. However, the process $H \to ZZ$ can still appear at one-loop level. Among the loop induced decays, $H \to gg$ is the dominant channel and the diphoton decay branching ratio $Br(H \to \gamma\gamma)$ can reach ${\cal O}(10^{-3})$. Due to the gauge symmetry, there is a strong correlation between the decay branching ratios of $H \to \gamma\gamma$, $H \to Z\gamma$ and $H \to ZZ$, which is $1:2s_W^2/c_W^2:s_W^4/c_W^4$ for small mixing angle.  When $|\sin\theta|>0$, the SM decay channels become accessible and then $Br(H \to \gamma\gamma)$ are severely suppressed. It should be noted that the branching ratio of $H \to hh$ becomes small when $\sin\theta$ is close to $-0.4$ since the Higgs self-coupling $c_{Hhh}= c_\theta (c_\theta^2-2s_\theta^2)\lambda_2 f_2+3 c_\theta^2s_\theta v \lambda_3$ will vanish at $\sin\theta \simeq -0.4$.

\begin{table}[t]
\begin{center}
\begin{tabular}{c|c|c} \hline
 Channel & CMS bound [fb] & ATLAS bound [fb] \\
\hline
	$gg$ & $1.8 \times 10^3$~\cite{jj1} & -- \\
	$ZZ$ & 27~\cite{zz1} & 12~\cite{zz2} \\
	$Z \gamma$ & -- & 6~\cite{zabound} \\
	$WW$ & 220~\cite{zz1} & 38~\cite{zz2} \\
	$hh$ & 52~\cite{hh2} & 35~\cite{hh2} \\
	$t\bar{t}$ & $6\times 10^2$~\cite{tt1} &  $7\times 10^2$~\cite{tt2}\\
	$\gamma\gamma$ & 1.3~\cite{aa1} & 2.2~\cite{aa2} \\
\hline
 \end{tabular}
\caption{\label{tab:constraints} 95\%~CL exclusion limits on $\sigma(g g \to H)\times Br(H\to X X)$ for various decay channels $X X$ in resonance searches at LHC Run-I.}
\end{center}
\end{table}

\begin{figure}[ht]
	\centering
	\includegraphics[width=0.32\textwidth]{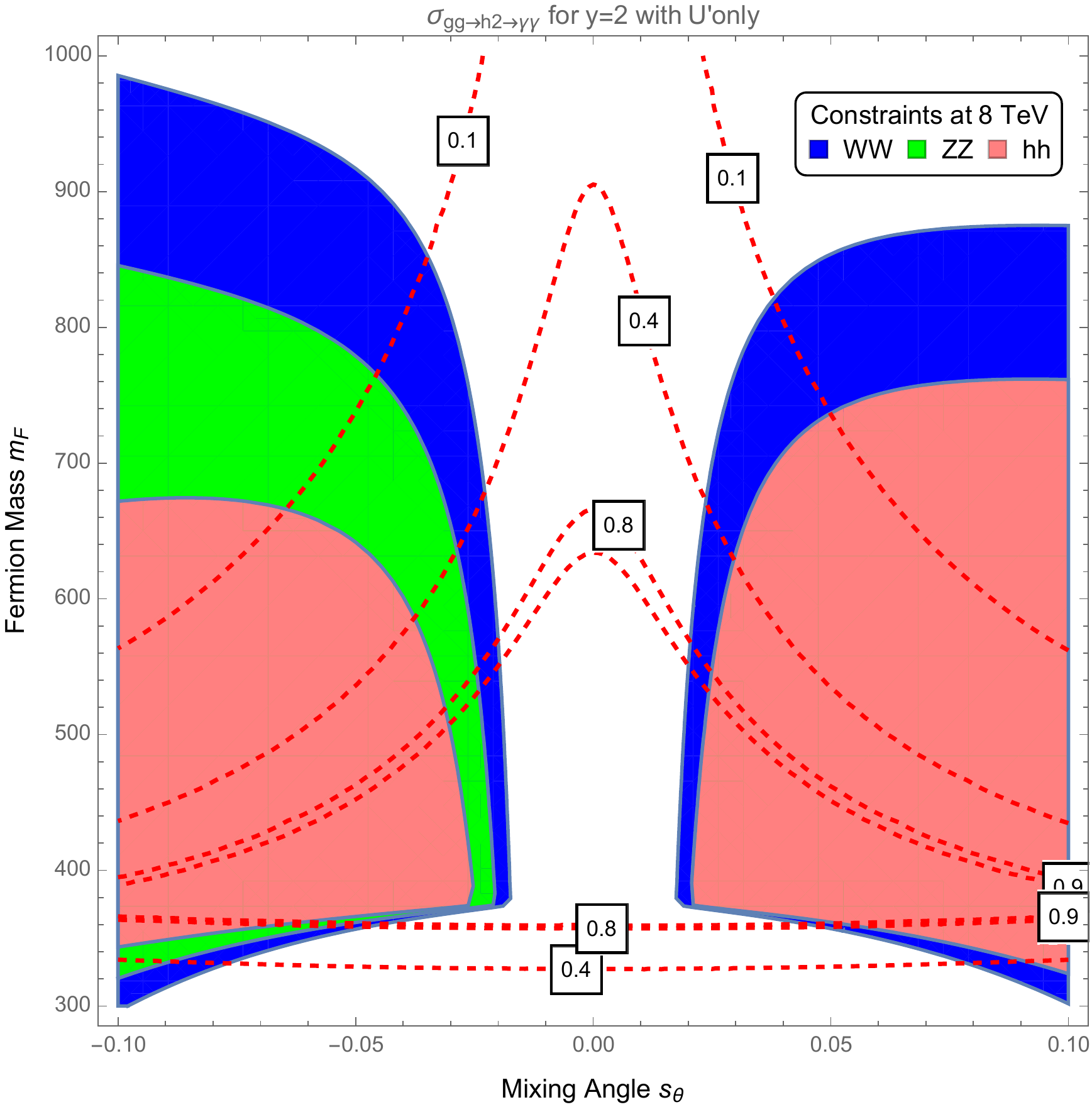}
	\includegraphics[width=0.32\textwidth]{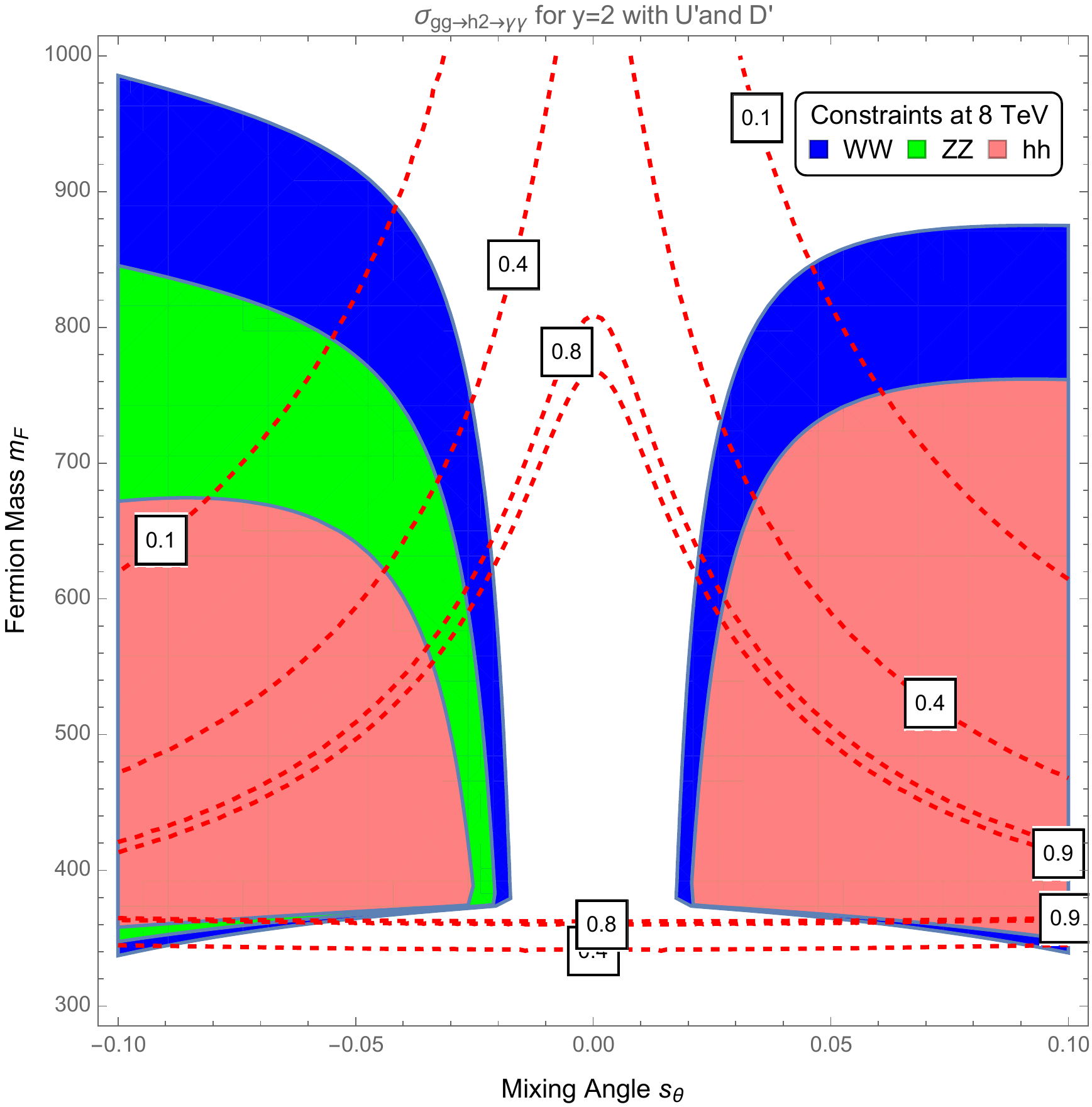}
	\includegraphics[width=0.32\textwidth]{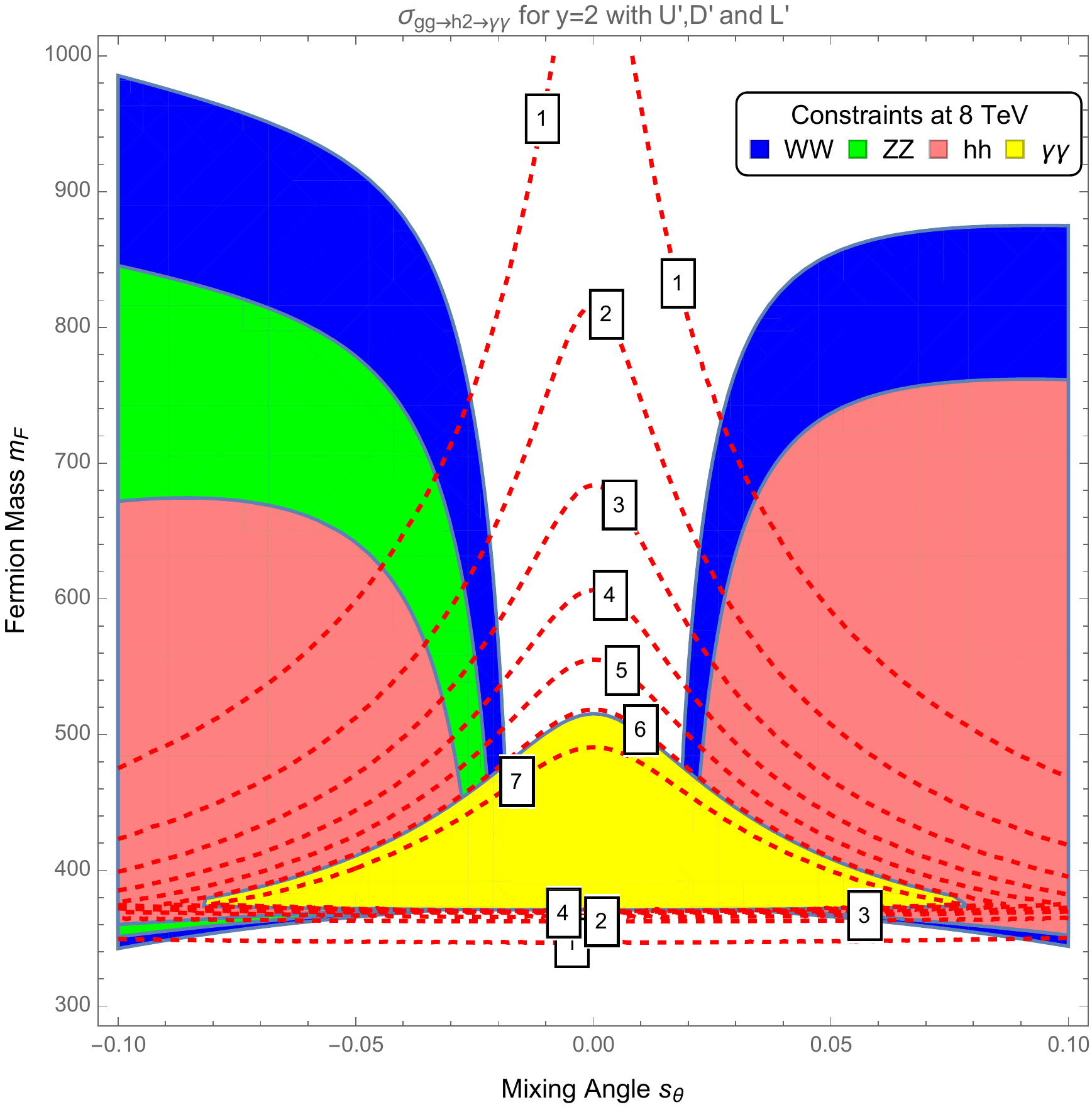}
	\caption{LHC constraints on the plane of $s_\theta$ versus $m_F$ for the cases $U'$, $U'+D'$ and $U'+D'+L'$ respectively. The Yukawa couplings $y_F$ are assumed as $y_F=2$, where $F=U',D',L'$. The red dashed contours correspond to the cross sections of $\sigma(g g \to H)\times Br(H\to \gamma\gamma)$ at 13 TeV LHC.}
	\label{fig:s.m4u}
\end{figure}

In Tab.\ref{tab:constraints}, we list the 95\%~CL exclusion limits on $\sigma(g g \to H)\times Br(H\to X X)$ for various decay channels in the resonance searches at LHC Run-I. Since no significant excesses were observed in these channels, in Fig.~\ref{fig:s.m4u}, we present these constraints on the plane of $s_\theta$ versus $m_F$ for the cases of $U'$, $U'+D'$ and $U'+D'+L'$, respectively. From Fig.~\ref{fig:s.m4u}, we can see that the most stringent constraint on the large mixing angle $s_\theta$ is from the measurement of the high mass resonance in $WW$ final states, which is then followed by $ZZ$ and $hh$ (other bounds from $jj$, $t\bar{t}$ and $Z\gamma$ are weak.). The 8 TeV diphoton measurement can exclude the small mixing angle region only for the case of $U'+D'+L'$. Depending on the sign of $s_\theta$, the amplitudes of the top quark and vector-like quarks can (de-)constructively interfere and lead to the asymmetric exclusion regions around $s_\theta=0$. When the mixing angle becomes larger, the vector-like fermions should be lighter to increase the diphoton production rate. However, if $m_F < 375$ GeV, the decay channel $H \to F\bar{F}$ can be accessible, then the diphoton decay branching ratio will be highly suppressed. It should be noted that only vector-like quarks can hardly enhance the diphoton production rate of $H$ to the observed value at 13 TeV LHC unless a very large Yukawa coupling. With the contribution of vector-like lepton, the diphoton production cross section can maximally reach 6 fb within the allowed parameter space.


\begin{figure}[h]
	\centering
	\includegraphics[width=10cm]{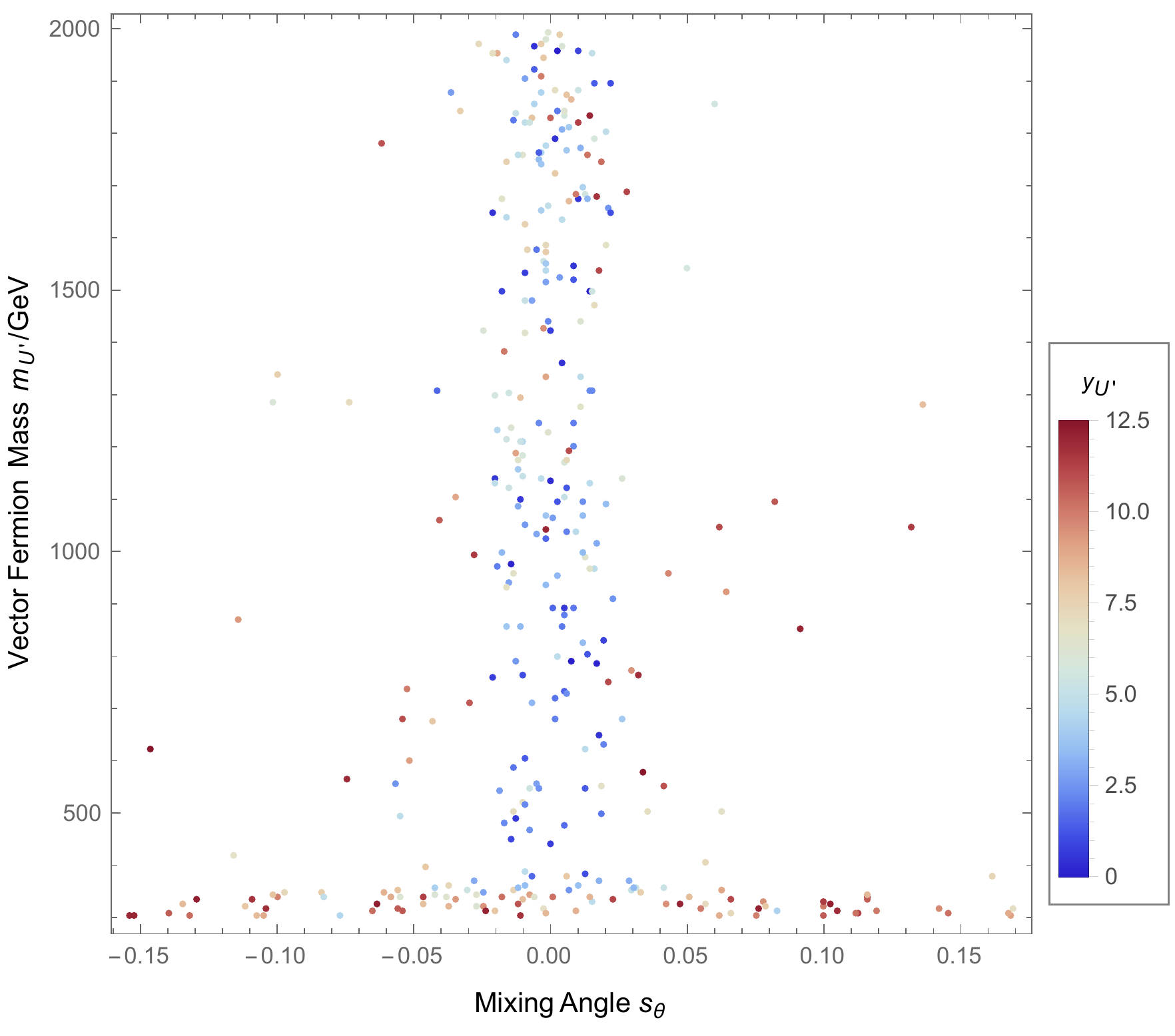}
	\caption{Scatter plots on the plane of $s_\theta$ versus $m_F$. All the samples satisfy the constraints in Tab.\ref{tab:constraints} and the $2\sigma$ range of Eq.(\ref{excess}). The color map corresponds to the Yukawa coupling $y_{U'}$.}
	\label{fig:scatter}
\end{figure}

In the above discussions, we have assumed universal vector-like fermion masses and Yukawa couplings. Now we will scan the full 7 dimensional parameter space $(m_{U'}, y_{U'}, m_{D'}, y_{D'}, m_{L'}, y_{L'}, s_\theta)$ within the ranges of $m_i\in[300,2000]$ GeV, $y_i\in [0,4\pi]$ and $s_\theta \in [-1,1]$. All the samples are required to satisfy the constraints in Tab.\ref{tab:constraints} and the $2\sigma$ range of Eq.(\ref{excess}). The surviving points are presented in Fig. \ref{fig:scatter}. It can be seen that the mixing angle is constrained within $|s_\theta| \lesssim 0.15$. This is because the large mixing angle will significantly suppress diphoton decay branching ratio. When $|s_\theta|$ approaches 0.15, the mass of $U'$ quark should be small ($\lesssim 500$ GeV) and the Yukawa coupling $y_{U'}$ has to become very large but is still bounded by the perturbativity requirement $y_{U'} < 4\pi$. While, if $|s_\theta| \rightarrow 0$, the mass of $U'$ quark can be heavier and the Yukawa coupling $y_{U'}$ is allowed to be smaller.

Besides the diphoton excess, other interesting signatures are predicted at the LHC and ILC in our model, such as the vector-like quark/lepton pair productions $pp/e^+e^- \to F \bar{F}$ ($F=U',D',L'$) \cite{vf} and the diboson productions $pp \to Z\gamma, ZZ, WW, hh$. With more data in LHC run-2, both ATLAS and CMS analyses will be able to confirm the 750 diphoton excess if it is indeed a signal of new physics beyond the SM. Then, these correlated signatures will be helpful to further test our model at the LHC and ILC.

\section{conclusions}\label{sec4}
In this work, we investigate the recent 750 GeV diphoton excess in the Hill Model augmented with the vector-like fermions. The heavy singlet-like Higgs is chosen as 750 GeV resonance and is mainly produced by the gluon fusion through vector-like top and bottom quarks. However, the mixing of singlet and doublet Higgs bosons suffers from the severe constraints from the searches for the resonance in dijet, $t\bar{t}$ and diboson channels. Under the current experimental and theoretical constraints, we find the viable parameter space that fits the 750 GeV diphoton signal strength at 13 TeV LHC. We note that the mixing angle of singlet and doublet Higgs boson should be $|\sin\theta| \lesssim 0.15$ to be consistent with the LHC constraints. The larger mixing angle becomes, the larger Yukawa couplings are required. In the allowed parameter space, the diphoton cross section can be maximally enhanced to about 6 fb at 13 TeV LHC. If the diphoton excess was further confirmed, other correlated signatures, such as the vector-like quark/lepton pair productions $pp/e^+e^- \to F \bar{F}$ ($F=U',D',L'$) and the diboson productions $pp \to Z\gamma, ZZ, WW, hh$, can be used to test our model at the LHC and ILC.

\acknowledgments
We thank Lei Wu for helpful discussions. This work is partly supported by the Australian Research Council, by the National Natural Science Foundation of China (NNSFC) under grants Nos. 11275057, 11305049, 11375001, 11405047, 11135003, 11275245, by Specialised Research Fund for the Doctoral Program of Higher Education under Grant No. 20134104120002, by the Startup Foundation for Doctors of Henan Normal University under contract No.11112.

\end{document}